\documentclass{ptapap}
\usepackage{color}
\definecolor{todo}{rgb}{0.8,0.2,0.3}
\author{Andrzej Pigulski}[IAUWr]
\author{Adam Popowicz}[Gliwice]
\author{Rainer Kuschnig}[Graz,UV]
\author{the BRITE Team}
\affil[IAUWr]{Instytut Astronomiczny, Uniwersytet Wroc{\l}awski, Wroc{\l}aw, Poland}
\affil[Gliwice]{Institute of Automatic Control, Silesian University of Technology, Gliwice, Poland}
\affil[Graz]{Graz University of Technology, Institute of Communication Networks and Satellite
Communications, Graz, Austria}
\affil[UV]{Institute of Astronomy, University of Vienna, Austria}
\title{Instrumental effects in BRITE photometry}
\begin{document}
\maketitle
\begin{abstract}
The raw photometry from BRITE satellites suffers from several instrumental effects. We present the list of the known effects and discuss their origin and the ways to correct for them.
\end{abstract}
\section{Introduction}
The BRITE-Constellation \citep{2014PASP..126..573W,2016PASP..128l5001P} is the first fleet of nano-satellites aimed at doing space photometry in the visual domain. They host wide-field (24$^\circ$ field of view) 3-cm refractors and uncooled CCD detectors. The stellar images are intentionally defocused to get FWHM\footnote{FWHM stands for the full width at half maximum.} of stellar profiles equal to several pixels. The scale in the plane of the BRITE detectors equals 27 arcsec per pixel \citep{2016PASP..128l5001P}, making the typical size of stellar profiles a few arc minutes wide. This defines the angular resolution in BRITE images. 

Due to their low mass ($\sim$7\,kg), the satellites are not perfectly stabilized. The actual quality of the stabilization is defined by the performance of the star-trackers. As explained by \cite{2014PASP..126..573W}, this performance is better for BTr, BLb, and BHr\footnote{The designation of BRITE satellites in this paper follows that in the other papers, namely: BRITE-Austria $=$ BAb, UniBRITE $=$ UBr, BRITE-Toronto $=$ BTr, BRITE-Montr\'eal (defunct) $=$ BMb, BRITE-Lem $=$ BLb, and BRITE-Heweliusz $=$ BHr.} than the Austrian satellites (BAb and UBr), which host star-trackers of a different design. The r.m.s.~scatter of the pointing of the satellites depends on many factors, including the number of bright stars in the observed field. Typically, it amounts to a few arc~min. Since star-tracker images are not downloaded, the satellite orientation at the onset of observations of a new field must be sometimes corrected using a trial-and-error procedure. In poor stellar fields the fine-pointing may fail --- this is the main reason why only one field far from the Galactic plane, the CetEri I\footnote{The fields are named after the constellation(s); see \cite{2017A&A...605A..26P} for the explanation.} field, was observed by BRITEs until now.

The BRITEs also suffer from an increasing number of hot pixels, which occur as a result of the damaging effect of cosmic-ray protons on the CCDs. The effect is lower in BTr and BHr, whose detectors have been shielded prior to their launches. The hot pixels exhibit meta-stable levels, the so-called random telegraph signal (RTS) phenomenon \citep{2016PASP..128l5001P}. As explained by these authors, the new mode of observing (chopping) was introduced for all BRITEs since mid-2015 to mitigate the effect of the increasing number of hot pixels and other instrumental effects.

Both the imperfect stability of the satellites and defects of the detectors make photometry with BRITE images a challenging task. The pipelines used to obtain the raw BRITE photometry from images obtained in both modes of observing, stare and chopping, are described in detail by \cite{2017A&A...605A..26P}. This paper also contains a detailed description of the format of the final photometry (Data Releases 2 to 5, DR2\,--\,DR5) and the explanation of subraster, setup, and other concepts used in the present paper. An interested reader is therefore directed to that paper for the details.

For the reasons explained above (and some others, which will be explained below), the raw BRITE photometry is affected by several instrumental effects. The main purpose of this paper is to let the users of BRITE photometry know about these effects and --- more importantly --- present  the ways to correct for them or at least explain in which situations they can occur. The paper summarizes our present knowledge of these effects.

\section{Instrumental effects}
\subsection{Correlations with position, CCD temperature and other parameters}
The raw BRITE photometry is {\sl aperture photometry} obtained with {\sl constant} aperture, either circular or thresholded\footnote{Thresholded aperture is the closed area, the shape of which depends on the observed stellar profile; see \cite{2017A&A...605A..26P} for the details of the definition and optimization of the thresholded apertures.}. This photometry is done in the presence of many effects such as (we mention only the most important): (i) blurring of stellar images due to imperfect pointing of the satellites, (ii) intra-pixel variability\footnote{The correction for intra-pixel variability is the only correction which is done by the pipeline; see \cite{2017A&A...605A..26P}, Sect.~3.3. However, this is done with the assumption of the same dependence on pixel fraction for all pixels in a given subraster. This means that only the first-order effects of the intra-pixel variability are accounted for by the pipeline.}, (iii) thermal changes of the detector and the optics, (iv) the lack of flat-fielding. In particular, the direct consequence of blurring is that going from frame to frame, we measure different (and in principle unknown) parts of the stellar flux.  Consequently, aperture photometry will be affected by the dependence of the flux on the position of a star in the subraster, temperature of the CCD and the optics, the amount of smearing, and other factors. {\sl The presence of these dependencies is the strongest instrumental effect in the BRITE photometry.} 

\begin{figure}
\centering
\includegraphics[width=0.9\textwidth]{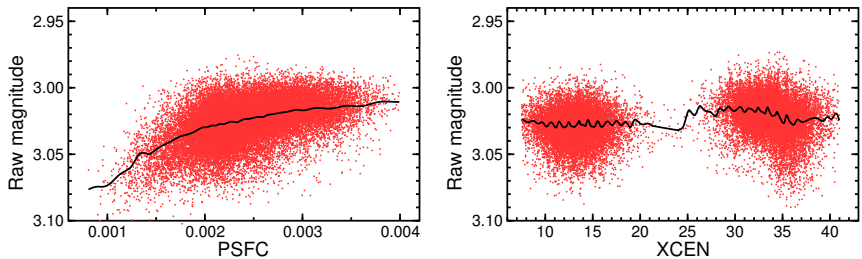}
\caption{An example of correlations between raw BRITE magnitudes and two parameters: the smearing parameter PSFC \citep[see][for its definition]{2017A&A...605A..26P} (left), and XCEN, the $x$ coordinate of the centroid of the stellar profile (right). Red dots are BRITE data. Black curves are functions which describe the correlations and are subtracted from the data during the decorrelation procedure. The data are HD\,186882 ($\delta$~Cyg) observations in Cygnus II field taken by the UBr satellite.}
\label{fig:decor}
\end{figure}

Fortunately, in most cases one can account for these dependencies by fitting them with a smooth function with subsequent subtraction. We call this procedure {\sl decorrelation}. The full discussion of the practical side of the procedure of decorrelation is beyond the scope of this paper --- some considerations related to this issue are presented in the `Cookbook 2.0' (Pigulski, these proceedings) and published papers \citep{2016A&A...588A..54W,2016A&A...588A..55P,2017A&A...602A..91B}. In practice, only one- (1D) and two-dimensional (2D) correlations are considered, although in general, higher-dimension decorrelations can be applied too. The 1D correlations are the strongest and at the beginning of our work with the BRITE data we thought that it is sufficient to account for these correlations only. An example of the correlation of the raw BRITE magnitudes with two parameters is shown in Fig.~\ref{fig:decor}. The wavy dependence on XCEN (right panel) is an illustration of the fact that in some cases the intra-pixel correction applied in the pipeline does not account for the real intra-pixel variability. In addition, a small magnitude offset between the two positions in the chopping mode can be seen in the same panel. The figure shows also that dependencies can hardly be approximated by a line or a higher-order polynomial. In a general case, more complicated functions (e.g.~splines) should be applied.

\begin{figure}
\centering
\includegraphics[width=0.85\textwidth]{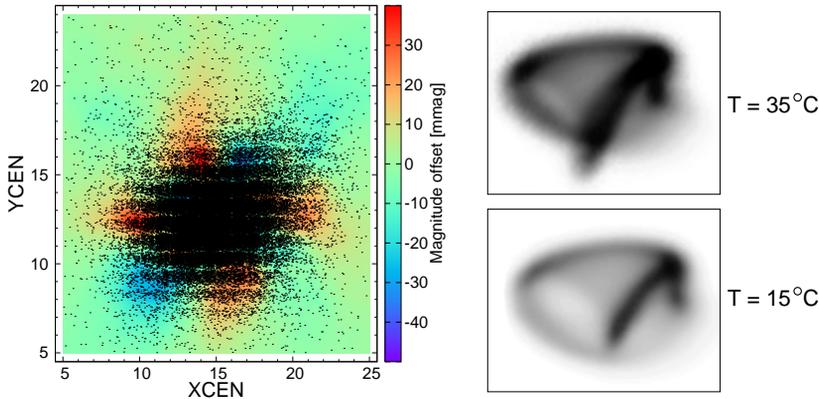}
\caption{Left: An example of 2D correlation, here between XCEN and YCEN, the $x$ and $y$ coordinates of the stellar centroid, respectively (HD\,195295 = 41~Cyg, BTr, Cyg\,I field). Right: Sample interpolated stellar image of the same star at two different temperatures.}
\label{fig:2d}
\end{figure}

The need for the 2D decorrelations was first recognized by Dr.\,Bram Buysschaert, who --- working with the BRITE data of $\zeta$~Ori --- found that the correlation of magnitude with one of the coordinates of the stellar profile centroid, changes with the temperature of the CCD. The effect was later explained by one of us (APo) as a consequence of the thermal expansion of optics and/or other elements of the BRITEs, which results in changes of the shape of the stellar profiles with the temperature. Figure \ref{fig:2d}, left panel, shows the colour-coded surface in magnitude that accounts for the 2D correlation between two parameters, XCEN and YCEN (explained in the caption of Fig.\,\ref{fig:decor}). This is the image after accounting for 1D decorrelations, so that it represents the residual effect. 2D correlations are usually much weaker than 1D ones. However, there are BRITE data in which the 2D effect is relatively strong and --- when 2D decorrelations are applied --- the reduction of scatter in the data is as much as 10 per cent.

The number of parameters provided with the BRITE data changed with the consecutive data releases; see Appendix A in the paper by \cite{2017A&A...605A..26P} for the description of parameters in different BRITE data releases. The only parameter which is measured at the satellite is CCDT, the temperature of the CCD detector. Another one, the satellite orbital phase, can be easily calculated. All the remaining parameters, including the position of the stellar centroid (XCEN, YCEN), are calculated from BRITE images by the reduction pipelines. They include parameters that are used to estimate the amount of smearing, the effect of the RTS phenomenon, and others. In the most recent data release (DR5), eight different parameters are provided. Adding calculated satellite orbital phase, the user has nine different parameters to decorrelate the data. This usually allows an efficient removal of most of the instrumental effects from the BRITE data.

\subsection{Saturation and non-linearities}
\begin{figure}
\centering
\includegraphics[width=\textwidth]{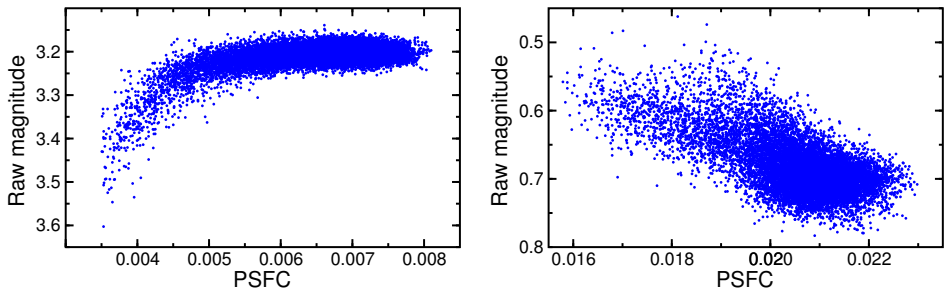}
\caption{A comparison of the effect of blurring on raw magnitudes of two stars, a non-saturated one (left panel; $\theta$~Oph, $V=$ 3.3~mag, BAb, setup 2, Sgr\,II field) and a saturated one (right panel; $\alpha$~Ori, $V=$ 0.1~mag, BAb, setup 2, Ori\,IV field). The PSFC parameter, defined by \cite{2017A&A...605A..26P}, is the measure of blurring --- the larger the blurring, the smaller the PSFC.}
\label{fig:nonl}
\end{figure}

The BRITEs observe stars with magnitudes down to $V\sim$ 7~mag (with the median value of $V=$ 4.2~mag, Pigulski, these proceedings) with practically no bright limit. In spite of defocussing, the signal in some pixels of the brightest stars falls into the non-linear regime of the CCD or even saturates. This regime starts approximately at the level of 9\,000~ADU and the CCD becomes highly non-linear above 12\,000~ADU \citep[][Sect.~3.2.3 and Fig.\,7]{2016PASP..128l5001P}. This is a consequence of the presence of the anti-blooming gates in the BRITE CCDs. That the non-linearity effect can be important for some BRITE targets was found by \cite{2016A&A...588A..55P} when analyzing BRITE data of $\beta$~Cen ($V=$ 0.6~mag), the 10th brightest star in the sky. A combination of blurring and constant aperture leads to the loss of a part of the measured flux. In general, the larger the blurring, the larger the loss of the flux. Therefore, for blurred images, that is, images with small values of the PSFC parameter\footnote{The PSFC parameter is a measure of the amount of blurring. The larger the blurring, the smaller the value of PSFC.}, the amount of the measured flux is smaller and the star looks fainter (Fig.\,\ref{fig:nonl}, left panel). For saturated stars, the effect is, paradoxically, the opposite (Fig.\,\ref{fig:nonl}, right panel). This can be explained if we realize that blurring cancels saturation and therefore {\sl increases} the measured signal in the centre of the stellar profile. If the increase is larger than the loss, the net effect of blurring is an increase of the measured flux like in the example shown in Fig.\,\ref{fig:nonl}, right panel.

While the non-linearity effect can be partly corrected by decorrelation with the smearing parameters, the measured amplitudes of periodic variation might be affected by the non-linearity. This is only a suspicion because there is no other source of precise photometry that can be presently used to verify this hypothesis and find the importance of the suspected effect.

It is difficult to provide the limiting magnitude for the non-linearity effect because the saturation depends on the shape of the stellar profile, filter, and exposure time. A short review of the raw photometry indicates that the non-linearity becomes relatively clearly visible for stars with $V<$ $\sim$1.5~mag unless shorter exposures were applied to avoid saturation. The sample of affected stars is therefore rather small, a dozen or so.

\subsection{CTI effect and CTI-related signal trapping effect}
\begin{figure}[!ht]
\centering
\includegraphics[width=0.7\textwidth]{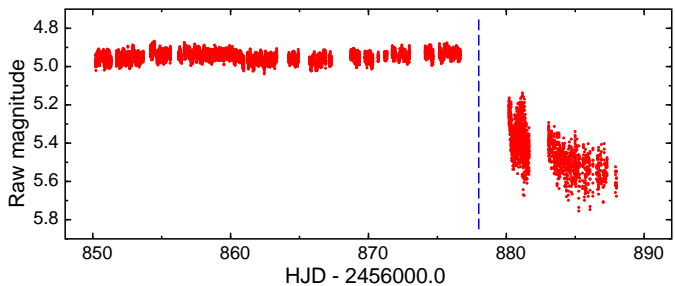}
\caption{An example of BRITE data affected by the occurrence of the CTI effect: BTr observations of HD\,191610 in the Cyg\,I field. The data made after $\sim$HJD\,2\,456\,878 (vertical dashed line) are affected by the CTI.}
\label{fig:cti}
\end{figure}

As explained in detail in Sect.~6 of the paper by \cite{2016PASP..128l5001P}, the charge transfer inefficiency (CTI) effect occurred in some areas of the CCDs of some BRITEs shortly after the onset of observations. The effect is a consequence of defects in the CCD structure caused by hits of cosmic-ray protons. These defects (`traps') result in the occurrence of streaked stellar images and consequently much poorer photometry. An example of the occurrence of CTI in the area of an observed star is shown in Fig.\,\ref{fig:cti}. After the gap in observations at $\sim$HJD\,2\,456\,878, there is a sudden drop in magnitude and an increase of scatter. This is a consequence of the occurrence of the CTI effect. There is practically no remedy to correct data strongly affected by CTI; usually, they are removed and lost.
\begin{figure}[!ht]
\centering
\includegraphics[width=0.63\textwidth]{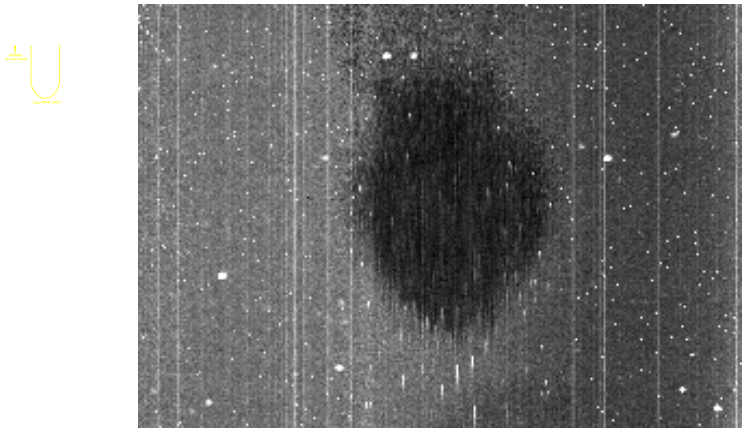}
\caption{An illustration of the CTI-related trapping effect in the BTr full-frame image exposed to stray light from Earth; see also Fig.~2 in the paper by Rucinski et al., these proceedings. The dark patch in the centre (roughly an ellipse of 90\,$\times$\,120 pixels) is the area from which the signal is stolen by traps. Its depth amounts to about 50~ADU. The smearing of the signal in hot pixels by CTI can be seen in the whole patch and downwards, in the direction of the serial register (the area encompassed by the dashed line).}
\label{fig:ctit}
\end{figure}

The effect of smearing by CTI was mitigated in July 2015 by slowing down the readout clocking rate for all BRITEs. This improved the photometry for stars located in the CTI areas, but did not remove the traps in the CCD. While analyzing the photometry of $\beta$~Lyr (CygLyr\,I field), another manifestation of the CTI was found (Rucinski et al., these proceedings, Sect.~3). A comparison of simultaneous UBr and BTr photometry of the star showed that some signal measured by BTr is lost. As a result, an apparent increase of the amplitude measured by this satellite could be seen. The affected areas have a form of elongated patches in the CCD (Fig.\,\ref{fig:ctit}). The amount of the charge which is `stolen' when a star falls within a patch is not always the same. In the patch shown in Fig.\,\ref{fig:ctit} it amounts to about 50~ADU. This CTI-related signal trapping effect seems to affect stronger the red-filter satellites (especially UBr and BTr). The reason for this behaviour is not known yet. The effect probably will become more severe with time because the number of patches likely grows with time. It remains to be investigated for which stars the photometry is affected and if the effect can be corrected. The rule-of-thumb remedy for this effect, applied by Rucinski et al.~(these proceedings), is to derive the lost flux by comparing the amplitudes from two satellites (BTr and UBr in the case of $\beta$~Lyr). This can be performed only if simultaneous data from two satellites hosting the same filter are available.

\subsection{Stray light and the Moon effect}\label{stray}
The BRITE photometry is affected by the stray light from the Earth, Moon, and Sun. The effect is not easy to estimate as the BRITE CCDs do not have mechanical shutters.  For observations obtained in the stare mode of observing, background due to the stray light is subtracted via the subtraction of column medians; see Sect.~3.2 of \cite{2017A&A...605A..26P}. However, a residual effect of the background can remain when it has a non-negligible gradient within a subraster. For the chopping mode of observing, rapidly varying background may affect photometry which is carried out using differential images. In order to account for this effect, the APER0 parameter has been introduced in DR5. For the earlier releases, this parameter was not provided and therefore decorrelated DR2\,--\,DR4 data can be more affected by stray light than DR5.
\begin{figure}
\centering
\includegraphics[width=0.7\textwidth]{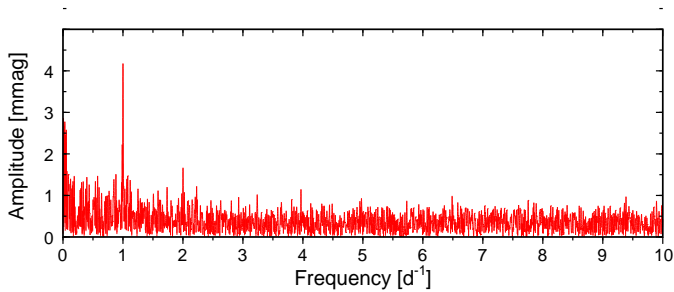}
\caption{Frequency spectrum of the UBr data of HD\,121743 ($\phi$ Cen; field Cen\,I), in which significant peaks at 1 and 2~d$^{-1}$ due to stray light from Earth can be seen.}
\label{fig:strl}
\end{figure}

The stray-light effect appears in the BRITE data in at least two different ways: (i) as a periodic signal at 1 and 2~d$^{-1}$; the origin is the stray light from Earth, (ii) as the `Moon effect', coming obviously from stray light due to Moon. It is also possible that some extra signal at the lowest frequencies is due to unaccounted-for stray light. An example of the first kind of the manifestation of the stray light is shown in Fig.\,\ref{fig:strl}. There has been no detailed study of the occurrence of the signal at 1 and 2~d$^{-1}$ in the frequency spectra of BRITE data, and in general it is not known why it appears in the BRITE data, but given the frequencies at which it occurs, stray light from the Earth is an obvious origin of this signal. Undoubtedly, the stray light from Earth is modulated with the satellite orbital period, but this contribution might be corrected for by a decorrelation with satellite orbital phase. The occurrence of 1 and 2~d$^{-1}$ signals has to be related to daily changes of the reflectance of some bright areas on the Earth. From the experience in working with BRITE data we can say that stare mode photometry is more affected by this effect than chopping mode photometry. In particular, the effect is unusually strong in the UBr and BAb Cen\,I data, in which amplitudes up to 6~mmag at 1~d$^{-1}$ are detected. In addition, the signal has a larger amplitude in the red-filter data than in the blue-filter data. It is also known that decorrelations, both 1D and 2D, lower these signals. The remedy is to subtract terms with frequencies at 1 and 2~d$^{-1}$ or ignore them as spurious.

\begin{figure}
\centering
\includegraphics[width=0.7\textwidth]{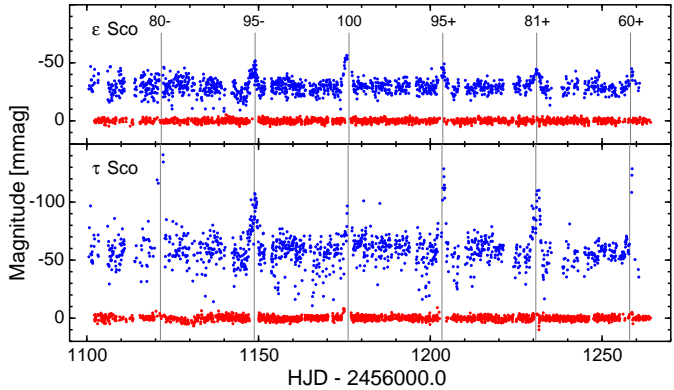}
\caption{Moon effect for two stars in the Sco I field, $\varepsilon$~Sco (top panel) and $\tau$~Sco (bottom panel). The dots are orbit averages for observations made by the red-filter UBr (red dots) and blue-filter BLb (blue dots) satellites. Vertical lines mark epochs of the smallest Moon elongation; the numbers are Moon phases in per cent (100 per cent is the full Moon). The sign indicates waxing (`$+$') or waning (`$-$') Moon.}
\label{fig:moon}
\end{figure}

The Moon effect is the most pronounced manifestation of the stray-light effect. It was discovered when the photometry of stars in the Sco~I field was analyzed. Later on, it was also found in the data from other fields, like Orion. Figure~\ref{fig:moon} shows an example for two stars located in the Sco~I field. The short brightenings coincide perfectly with the epochs of the smallest Moon elongation, which amounted to about 16$^\circ$ for $\varepsilon$~Sco and only 10$^\circ$ for $\tau$~Sco. This is why the effect is stronger for $\tau$~Sco. The effect is stronger in blue than in the red filter, although this conclusion is uncertain because UBr observations were suspended when the Moon was close to the observed field --- this is the reason for small gaps in the UBr data.

The effect was later found for some stars in the Ori field and can be expected for other BRITE fields having small ecliptic latitudes. In addition to the Sco and Ori fields, this refers mainly to the Sgr and AurPer fields and the Tau field, scheduled for observation this year. The effect is hard to correct, so that the obvious remedy seems to be a removal of the affected data, typically an interval of 1\,--\,2 days.

\begin{figure}
\centering
\includegraphics[width=0.57\textwidth]{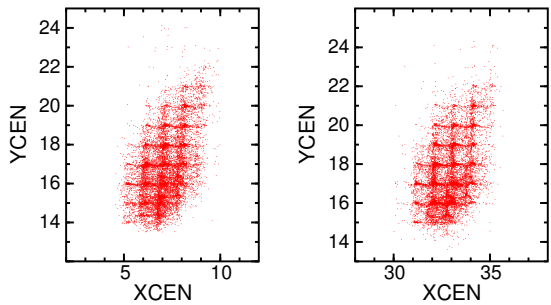}
\caption{Positions of stellar centroids (XCEN,YCEN) as derived by the photometric pipeline for part of BHr observations of HD\,50877 ($o^1$~CMa) in the CMaPup\,I BRITE field. The two panels correspond to the two positions of the star in the subraster when observed in the chopping mode.}
\label{fig:grid}
\end{figure}

\subsection{Tracker gridding}
While working with the BHr data for stars in the CMaPup\,I field, it was found that the positions of stellar centroids do not cluster around a unique position with some scatter, but tend to be concentrated along a well-defined grid (Fig.\,\ref{fig:grid}), with the grid eye size similar to the pixel size of the BRITE CCD. (Another example of gridding in positions can be seen in Fig.\,\ref{fig:2d}, but this time the positions along the grid are avoided.) This led to the suspicion that the procedure of centroid determination might be responsible for the gridding. The suspicion appeared to be wrong because this kind of behaviour was not found for all observations. Detailed analysis of the temporal behaviour allowed to explain the phenomenon as due to the way in which the BRITE trackers work and correct the positions of the satellites. Trackers in BRITEs work in their own loops and, at least for BHr, BLb, and BTr, have virtually the same pixel scale as the main CCDs. 

The tracker gridding has no direct influence on the photometry and needs no correction. The only difficulty related to this effect is that the data are highly non-evenly distributed in the space of the XCEN and YCEN parameters. This may slightly complicate decorrelations.

\section{Conclusions}
The instrumental effects in BRITE data are relatively well understood and we know how to correct for them. The only exception is the CTI-related signal trapping effect, which will be investigated in the near future as it may become the most severe problem deteriorating the forthcoming BRITE observations (the number of CTI-affected patches likely grows with time). Excluding the occurrence of weak 1~d$^{-1}$ and 2~d$^{-1}$ signals, no periodic instrumental features were detected in BRITE data. In some data, small increase of noise towards low frequencies and an excess power at frequencies below 0.1~d$^{-1}$ can be seen (Fig.\,\ref{fig:strl} shows an example), but these features disappear in most cases in which a large set of decorrelation parameters is used and 1D and 2D decorrelations are performed. Finally, in some stars with long-term intrinsic variability, the separation of the intrinsic variability and instrumental effects may be ambiguous. The reason is that some parameters, like CCDT, may also change on a long timescale.

\acknowledgements{The study is based on data collected by the BRITE Constellation satellite mission, designed, built, launched, operated and supported by the Austrian Research Promotion Agency (FFG), the University of Vienna, the Technical University of Graz, the Canadian Space Agency (CSA), the University of Toronto Institute for Aerospace Studies (UTIAS), the Foundation for Polish Science \& Technology (FNiTP MNiSW), and National Science Centre (NCN). The operation of the Polish BRITE satellites is secured by a SPUB grant of the Polish Ministry of Science and Higher Education (MNiSW). APi and APo acknowledge support from the NCN grants no.~2016/21/B/ST9/01126 and 2016/21/D/ST9/00656, respectively.}
\bibliographystyle{ptapap}

\end{document}